\newcommand{\be}{\begin{equation}}\newcommand{\ee}{\end{equation}}
\newcommand{\bea}{\begin{eqnarray}}\newcommand{\eea}{\end{eqnarray}}
\documentstyle[12pt]{article}
\def\PRL #1 #2 #3{{\sl Phys. Rev. Lett.} {\bf#1} (#2) #3}
\def\NPB #1 #2 #3{{\sl Nucl. Phys.} {\bf B#1} (#2) #3}
\def\NPBFS #1 #2 #3 #4{{\sl Nucl. Phys.} {\bf B#2} [FS#1] (#3) #4}
\def\CMP #1 #2 #3{{\sl Commun. Math. Phys.} {\bf #1} (#2) #3}
\def\PRD #1 #2 #3{{\sl Phys. Rev.} {\bf D#1} (#2) #3}
\def\PLA #1 #2 #3{{\sl Phys. Lett.} {\bf #1A} (#2) #3}
\def\PLB #1 #2 #3{{\sl Phys. Lett.} {\bf #1B} (#2) #3}
\def\JMP #1 #2 #3{{\sl J. Math. Phys.} {\bf #1} (#2) #3}
\def\PTP #1 #2 #3{{\sl Prog. Theor. Phys.} {\bf #1} (#2) #3}
\def\SPTP #1 #2 #3{{\sl Suppl. Prog. Theor. Phys.} {\bf #1} (#2) #3}
\def\AoP #1 #2 #3{{\sl Ann. of Phys.} {\bf #1} (#2) #3}
\def\PNAS #1 #2 #3{{\sl Proc. Natl. Acad. Sci. USA} {\bf #1} (#2) #3}
\def\RMP #1 #2 #3{{\sl Rev. Mod. Phys.} {\bf #1} (#2) #3}
\def\PR #1 #2 #3{{\sl Phys. Reports} {\bf #1} (#2) #3}
\def\AoM #1 #2 #3{{\sl Ann. of Math.} {\bf #1} (#2) #3}
\def\UMN #1 #2 #3{{\sl Usp. Mat. Nauk} {\bf #1} (#2) #3}
\def\FAP #1 #2 #3{{\sl Funkt. Anal. Prilozheniya} {\bf #1} (#2) #3}
\def\FAaIA #1 #2 #3{{\sl Functional Analysis and Its Application} {\bf
#1} (#2) #3}
\def\BAMS #1 #2 #3{{\sl Bull. Am. Math. Soc.} {\bf #1} (#2)
#3} \def\TAMS #1 #2 #3{{\sl Trans. Am. Math. Soc.} {\bf #1} (#2) #3}
\def\InvM #1 #2 #3{{\sl Invent. Math.} {\bf #1} (#2) #3}
\def\LMP #1 #2 #3{{\sl Letters in Math. Phys.} {\bf #1} (#2) #3}
\def\IJMPA #1 #2 #3{{\sl Int. J. Mod. Phys.} {\bf A#1} (#2) #3}
\def\AdM #1 #2 #3{{\sl Advances in Math.} {\bf #1} (#2) #3}
\def\RMaP #1 #2 #3{{\sl Reports on Math. Phys.} {\bf #1} (#2) #3}
\def\IJM #1 #2 #3{{\sl Ill. J. Math.} {\bf #1} (#2) #3}
\def\APP #1 #2 #3{{\sl Acta Phys. Polon.} {\bf #1} (#2) #3}
\def\TMP #1 #2 #3{{\sl Theor. Mat. Phys.} {\bf #1} (#2) #3}
\def\JPA #1 #2 #3{{\sl J. Physics} {\bf A#1} (#2) #3}
\def\JSM #1 #2 #3{{\sl J. Soviet Math.} {\bf #1} (#2) #3}
\def\MPLA #1 #2 #3{{\sl Mod. Phys. Lett.} {\bf A#1} (#2) #3}
\def\JETP #1 #2 #3{{\sl Sov. Phys. JETP} {\bf #1} (#2) #3}
\def\JETPL #1 #2 #3{{\sl  Sov. Phys. JETP Lett.} {\bf #1} (#2) #3}
\def\PHSA #1 #2 #3{{\sl Physica} {\bf A#1} (#2) #3}
\def\CQG #1 #2 #3{{\sl Class. Quantum Grav.} {\bf #1} (#2) #3}
\def\a{\alpha}\def\b{\beta}\def\g{\gamma}\def\d{\delta}\def\e{\epsilon}
\def\k{\kappa}\def\s{\sigma}
\def\Th{\Theta}\def\th{\theta}\def\om{\omega}\def\G{\Gamma}
\def\ws{worldsheet}
\def\susy{supersymmetry}

\def\ks{$\k$--symmetry}

\newcommand{\nn}{\nonumber\\}\newcommand{\p}[1]{(\ref{#1})}
\begin{document}
\renewcommand{\thefootnote}{\fnsymbol{footnote}}
\thispagestyle{empty}
\begin{center}
{\large\bf $N=1$ SUPER--$P$--BRANES\\
in twistor-like Lorentz harmonic formulation}
\footnote{ The work was supported  in  part  by  the
American Physical Society.}
\footnote{ The work was supported  in  part  by  the Fund  for  Fundamental
Researches of the State  Committee  for  Science  and  Technology  of
Ukraine under the Grant N2/100.}
\vspace{1cm}
\\Igor A. Bandos
\footnote{ and {\it Ististuto Nazionale di Fisica Nucleare --
Sezione di Padova, Padova, Italy}.
//
E-mail up to June 8: bandos@pd.infn.it}
and Aleksandr A. Zheltukhin
\vspace{1cm}\\
\renewcommand{\thefootnote}{\dagger}
{\it Kharkov Institute of Physics and Technology}\\
{\it 310108, Kharkov, the Ukraine}\\
e-mail:  kfti@kfti.kharkov.ua\\

\vspace{1.5cm}
{\bf Abstract}
\end{center}

Unique twistor--like Lorentz harmonic formulation for all
$N=1$
supersymmetric extended objects
(super--$p$--branes) moving in the space--time of arbitrary dimension D
(admissible for given $p$) are suggested. The equations of motion are
derived, explicit form of the \ks{} transformations is presented and
the classical equivalence to the standard formulation is proved.
The cases
with minimal world--sheet dimensions $p=1,2$, namely of
$D=10$ heterotic string and $D=11$ supermembrane,
are
considered in details. In particular, the explicit form of irreducible
\ks{} transformations for $D=11$ supermembrane is derived.

\vspace{1cm}\\
{\bf DFPD / 94 /TH/35 ; hep-th / 9405113 }
\bigskip \\
{\bf May 1994}
\setcounter{page}1
\renewcommand{\thefootnote}{\arabic{footnote}}
\setcounter{footnote}0
\newpage
\section{Introduction}

The extension of different variants of the twistor approach \cite{pen}
for the case of supersymmetric extended objects (super--$p$--branes)
moving in high dimensional space-time became an actual task after the
works \cite{vz,stv,stvz} had been published. In
\cite{stv,stvz} the famous \ks{} of $D=3,4$ superparticle
\cite{al,sig}  have been identified with the world-line \susy.
Moreover, it has been proved \cite{stv}
that the problem of
irreducible \ks{} description which troubled superparticle and
superstring covariant quantization \cite{gsw} are automatically
solved in the framework of twistor-like formulation, i.e., when
an appropriate auxiliary bosonic spinor variables are present in
the configurational space of the theory and the action functional
has twistor-like form
\footnote{See also the papers \cite{twbc,twes,twplyus},
were the twistor formulations of supersymmetric particles and
strings are discussed}.

One of the directions of the generalizations of the results from Refs.
\cite{vz,stv,stvz} consist in the construction of world-line
(world-sheet) superfield or doubly supersymmetric formulation of the
supersymmetric objects, where the \ks{} is completely realized as
superconformal world-line (\ws) \susy{}. Such formulation have been
presented for superparticle in $D=3,4,6,10$ \cite{sp}, heterotic
\cite{hsstr} and $D=3$ Green-Schwarz superstrings \cite{gs93},
as well as for the supermembrane in $D=11$ \cite{ton}.

At the same time, a component variant of twistor-like approach has
been developed using the Lorentz harmonic variables as twistor-like
ones \cite{bh,bz0,bzst,bzm}. The works \cite{bh,bz0,bzst,bzm} gave the
bridge between the twistor approach and the well known works \cite{nis,kr},
where the idea of Lorentz covariant quantization of $D=10$ superparticle
and superstring using the extension of the phase space
by specially chosen set of vector \cite{sok} and spinor harmonics had been
realized.  Extended objects without tension (null super--$p$--branes) in
$D=4$ space-time have been covariantly quantized in the
twistor-like Lorentz harmonic formulation \cite{bz0}. This demonstrated
the power of such approach and gave the first example of selfconsistent
quantum theory for the extended objects with the world-volume dimension
$d=p+1 > 2$. The formulations of such type have been constructed for
$D=4, N=1$ and $D=10, N=IIB$ superstrings \cite{bzst} as well as
for  $D=11$ supermembrane \cite{bzm}.

Another super--$p$--branes \cite{m0} -- \cite{m3}
also are physically interesting objects,
because they appears as supersymmetric solitons in some field
theories \cite{m0,m1a} as well as in some super--$p$--brane
theories \cite{m1c,m1d}. So, the heterotic $5$--brane \cite{m1b} appears
as soliton solution in $D=10$ superstring theory \cite{m1c} and vice versa
\cite{m1d}.

Here we will suggest an universal twistor--like Lorentz-harmonic
formulation for the all admissible set \cite{t1}
of $N=1$ super-$p$-branes in
D-dimensional space-time. We will prove their classical
equivalence to the standard formulations, derive the motion
equations and present an explicit form of the
{\sl irreducible} \ks{}
transformations.

For simplicity, we give
an explicit calculations for the dimensions $D=2,3,4,10,11$ where
the Majorana spinors exist. The intermediate formulas for the
general case may be easily reproduced using slightly more complicated
notations of Ref.\cite{t1}.

\bigskip

\section{Twistor-like action, equations of motion and \ks{} for $N=1$
super--$p$--branes }

\bigskip

\begin{center}
{\sl \large {2.1. Lorentz harmonic variables.
\\ Definitions and admissible variation
concept}}
\end{center}

\bigskip
\bigskip

In this subsection we describe the necessary set of the Lorentz
harmonic variables which are suitable for the construction
of the twistor like formulations of $N=1$ super-p-branes
living in D dimensions for all admissible values $D$ and
$p$ \cite{m,t1}. For simplicity, explicit expressions will be given
for the cases $D=2,3,4(mod$  $8)$, where the Majorana
spinor representation exists. Of course, all the results may be
easily extended to another values of $D$ using the spinor conventions
 developed in \cite{t1}.

Hence we suggest that the used $\g$-matrices and $\s$-matrices
are symmetric and real ones

\begin{equation}\label{1.01}
(C\Gamma _{{ m}})^{{ T}} = (C\Gamma _{{ m}}),\qquad (\Gamma ^{{ n}}C^{-{
1}})^{{ T}} = (\Gamma ^{{ n}}C^{-{ 1}})
\end{equation}

Moreover, for the all admissible values of $D$ and $p$
(i.e., for all $D$ and $p$ where  the  standard $N=1$ super$-p$-brane
formulations exist \cite{m} )

\begin{equation}\label{1.02}
(\Gamma ^{{ m}_{{ 1}}}C^{-{ 1}})_{\{\alpha
\beta } (\Gamma _{{ m}_{{ 1}}{\bf ...}{ m}_{{ p}}} C^{-{ 1}})_{\gamma
\delta \}} = 0
\end{equation}

The Lorentz harmonic variables form the matrix $v^{{a}}_{\a}$ taking its
values in the spinor representation of the double covering of
D-dimensional Lorentz group ${\it Spin}(1,9)\simeq SO(1,9)$
\begin{equation}\label{1.1}
v^{{a}}_{\a} \in \hbox{ {\it Spin}}(1,D-1)
\end{equation}
In \p{1.1}
$\a=1,\ldots,2^{\nu}$ is $Spin(1,D-1) (\simeq SO(1,D-1))$
spinor index ;
$a=1,\ldots ,2^{\nu }$
is the composed spinor index of the right product
of (pseudo)orthogonal groups $[SO(1,p)\times SO(D-p-1)]$ and $2^\nu$ is
the dimension of the considered spinor representation: $\nu = [D/2]$ for
$D=3,4(mod{ }8)$ and $\nu = (D-2)/2$ for $D=2(mod{ }8)$ when
the Majorana-Weyl spinors are used.

The key point of the discussed approach consists in the statement
that the basic elements of the orthonormal vector repere
$u^{(n)}_{{ m}}$
\begin{equation}\label{1.2}
u^{{ (n)}}_{{ m}} u^{{ m(l)}}= \eta ^{(n)(l)}=
{\it diag}(1,-1,\ldots ,-1)
\end{equation}
are determined by means of the "square root" type
universal relations in terms of a harmonic variable matrix
$v^{{ a}}_{\alpha }$
\begin{equation}\label{1.3}
u^{{ (n)}}_{{ m}}\equiv 2^{-\nu } v^{{ a}}_{\alpha }
(C\Gamma _{{ m}})^{\alpha \beta } v^{{ b}}_{\beta }
(\Gamma ^{{ (n)}}C^{-{ 1}})_{{ ab}}
\end{equation}

For $D=2(mod{ }8)$ and the Majorana-Weyl spinors the matrices
$(C\Gamma _{{ m}})$ and
$(\Gamma ^{{n}}C^{-{ 1}})$ should be understood as
the chiral $\s$--matrices
$\tilde{\s}_{{ m}}$ and $\s^{{ n}}$
\begin{equation}\label{1.4}
(C\Gamma _{{ m}})
\rightarrow
\tilde{\s}_{{ m}} ,
(\Gamma ^{{n}}C^{-{ 1}})
\rightarrow
\s^{{ n}}
\end{equation}
Then, more strong relation is satisfied for $D=10, p=1$
instead of \p{1.02}
\begin{equation}\label{1.03}
\tilde{\sigma }^{\alpha
\{\beta }_{{ m}} \tilde{\sigma }^{\gamma \delta \}{ m}} \equiv  {{
1}\over { 3}} ( \tilde{\sigma }^{\alpha \beta }_{{ m}} \tilde{\sigma
}^{\gamma \delta { m}} +\hbox{ {\it cyclic permutations} }
(\alpha,\beta ,\gamma ))=0
\end{equation}

Eq.  \p{1.1} is realized as the requirement for the harmonic
matrix  to satisfy some algebraic restrictions which are called
harmonicity conditions
(see \cite{gikos})
\begin{equation}\label{1.1a}
v^{{ a}}_{\alpha } \in \hbox{
{\it Spin}}(1,D-1) \Leftrightarrow
\Xi _{M}(v) = 0
\end{equation}

An explicit form of the harmonicity conditions have been presented in
\cite{bh}, \cite{ghs,gds} for $D=3,4,6,10$ superparticles and superstrings
(see also \cite{zup} for $D=10$ case)
and in \cite{bzm} for $D=11$ supermembrane. Here we shell present only the
necessary set of general features of harmonic variables and will not
write these harmonicity conditions explicitly
\footnote{Such approach to harmonic variables was used in our previous
works \cite{bzm} as well as in the recent work \cite{ght}.
Previously the
similar approach to the spinor realization of the repere variables
(which are identical with Lorentz harmonic variables as well as
with generalized Newman--Penrose dyades \cite{bz0,bzst,bzm})
was used in the interesting work of Wiegmann \cite{wieg} devoted to the
effective actions of spinning and heterotic strings}.

As the consequence of \p{1.1a} the
$\Gamma$--matrices can be removed from one side of \p{1.3} to another.
I.e. there are the following consequences of \p{1.1} for any
$D=2,3,4(mod{}8)$
$$
\Xi_{M}(v) = 0
\Rightarrow
$$
\begin{eqnarray}\label{1.5}
u^{{ (n)}}_{{ m}} (\Gamma ^{{ m}}C^{-{ 1}})_{\alpha \beta } =
v^{{ a}}_{\alpha } (\Gamma ^{{ (n)}}C^{-{ 1}})_{{ ab}}
[5~v^{{ b}}_{\beta } \nn
u^{{ (n)}}_{{ m}} (C\Gamma _{{ (n)}})^{{ ab}} =
v^{{ a}}_{\alpha } (C\Gamma _{{ m}})^{\alpha \beta }
v^{{ b}}_{\beta }
\end{eqnarray}
\begin{eqnarray}\label{1.6}
(\Gamma ^{{ m}}C^{-{ 1}})_{\alpha \beta } =
v^{{ a}}_{\alpha } (\Gamma ^{{ (n)}}C^{-{ 1}})_{{ ab}}
v^{{ b}}_{\beta }  u^{{ m}}_{{ (n)}},
\nn
(C\Gamma _{{(n)}})^{{ ab}} =
v^{{ a}}_{\alpha } (C\Gamma _{{ m}})^{\alpha \beta }
v^{{ b}}_{\beta } u^{{ m}}_{{ (n)}},
\end{eqnarray}
\begin{eqnarray}\label{1.7}
u^{{ (n)}}_{{ m}} (\Gamma _{{ (n)}}C^{-{ 1}})_{{ ab}} =
v^{\alpha }_{{ a}} (\Gamma _{{ m}}C^{-{ 1}})_{\alpha \beta }
v^{\beta }_{{ b}}, \nn
u^{{ (n)}}_{{ m}} (C\Gamma ^{{ m}})^{\alpha \beta } =
v^{\alpha }_{{ a}} (C\Gamma _{{ (n)}})^{{ ab}} v^{\beta }_{{ b}}
\end{eqnarray}

Here $v^{\a}_{{a}} \equiv  v^{-{ 1 \alpha }}_{{ a}}$ ,
i.e. the relations
\begin{equation}\label{1.8}
\Xi ^{{ b}}_{{ a}} \equiv  v^{\alpha }_{{ a}} v^{{ b}}_{\alpha } -
\delta ^{{ b}}_{{ a}} = 0
\end{equation}
should be satisfied. This is an independent harmonicity condition for
$D=2(mod{}8)$. However, the matrix $v^{\alpha }_{{ a}}$ can be
constructed from $v^{{ b}}_{\alpha }$  one for $D=3,4(mod{ }8)$ due to
harmonicity conditions

\begin{equation}\label{1.8a}
\Xi ^{{ ab}} \equiv
v^{{ a}}_{\alpha } C^{\alpha \beta }v^{{ b}}_{\beta }
- C^{{ ab}} = 0 \Rightarrow  v^{\alpha }_{{ a}} =
C^{-{ 1}}_{{ ab}} v^{{ b}}_{\beta } C^{\beta \alpha }
\end{equation}
Eq.\p{1.8a} is the manifestation of the invariance of the charge
conjugation matrix $C^{{ ab}}$ under $SO(1,D-1)$ rotations.

Moreover, the following relations
\begin{eqnarray}\label{1.9}
Sp (v^{{ T}} C\Gamma _{{ m}_{{1}}{\bf ...}{ m}_{{ k}}} v
\Gamma ^{{ (n)}}C^{-{ 1}}) = 0 ,\qquad
Sp (v^{{ T}} C\Gamma _{{ m}} v
\Gamma ^{{ (n}_{{1}}{ )}{\bf ...}{ (n}_{{k}}{ )}}C^{-{ 1}}) = 0 , \nn
(\hbox{ {\it when} }k>1)  \hskip48pt  \qquad
\end{eqnarray}
are satisfied for the matrix
$v^{{a}}_{\a} \in {\it Spin}(1,D-1)$ \p{1.1}.

The relations \p{1.3}, \p{1.5} -- \p{1.7} are
{\sl the basic ones of the twistor--like Lorentz
 harmonic approach to super$-p$-brane theories}.

To solve the variational problem formulated in the configurational
space which includes the Lorentz-harmonic variables the concept of
admissible variations \cite{sok,bzst} is very useful. These are the
variations which do not violate the harmonicity conditions \p{1.1a}
or, equivalently, the relation \p{1.1}
\begin{equation}\label{1.10}
(v^{{a}}_{\alpha } + \delta v^{{ a}}_{\alpha })
\in \hbox{ {\it Spin}}(1,D-1)
\end{equation}

For the definition of the admissible variation it is convenient to discuss
the variation of the \p{1.5}--\p{1.7} arising from the
harmonicity conditions. So, varying the relation \p{1.5}, we get

\begin{eqnarray}\label{1.11}
\delta u^{{ (n)}}_{{ m}} (C\Gamma _{{ (n)}})^{{ ab}}
= \delta v^{{ a}}_{\alpha } (C\Gamma _{{ m}})^{\alpha \beta }
v^{{ b}}_{\beta } + v^{{ a}}_{\alpha }
(C\Gamma _{{ m}})^{\alpha \beta }
\delta v^{{ b}}_{\beta } \equiv \nn \nn
\equiv  (v^{-{ 1}}\delta v)^{{ a}}_{{ d}}
u^{{ m(k)}}(C\Gamma _{{ (k)}})^{{ db}} + (v^{-{ 1}}
\delta v)^{{ b}}_{{ d}} u^{{ m(k)}}(C\Gamma _{{ (k)}})^{{ da}}
\end{eqnarray}

Use of the first relation \p{1.7} transforms Eq.  \p{1.11}
into  the form

\begin{equation}\label{1.12}
u^{{ m(k)}}\delta u^{{ (n)}}_{{ m}}
(\Gamma _{{(k)(n)}})^{{ a}}_{{ b}} =
D (v^{-{ 1}}\delta v)^{{ a}}_{{ b}}+
(C\Gamma ^{{ (k)}}(v^{-{ 1}}\delta v)
\Gamma _{{ (k)}}C^{-{ 1}})^{{ a}}_{{ b}} \qquad
\end{equation}

Taking into account the Fiertz identities \p{A.1} (see Appendix A) for
the matrix $(v^{-{ 1}}\delta v)$, we can derive from \p{1.12} the
following set of  relations which defines the admissible variation

\begin{equation}\label{1.13}
Sp(v^{-{ 1}}\delta v) = 0
\end{equation}
\begin{equation}\label{1.14}
 - 2^{-{ (\nu -1)}} Sp( v^{-{ 1}}\delta v\Gamma ^{{ (k)(l)}})
= u^{{ m(k)}}\delta u^{{ (l)}}_{{ m}}
\equiv \Omega ^{{ (k)(l)}}(\delta )
\end{equation}
\begin{equation}\label{1.15}
Sp(v^{-{ 1}}\delta v \Gamma _{{ m}_{{ 1}}{\bf ...}{ m}_{{ q}}} ) = 0 ,
\end{equation}
where $q=1,3,4,\ldots $ for $D=3,4(mod$  $8)$,
and $q= 4,6,\ldots $ for $D=2(mod$  $8)$.

Hence, the admissible variation of $v^{{a}}_{{\a}}$ is one which
may be reduced to $SO(1,D-1)$ rotation
\begin{equation}\label{1.16}
\delta v^{{a}}_{{\a}} =
{1\over 4}
\Omega ^{{ (k)(l)}}(\delta )
(\Gamma _{{ (k)(l)}})^{{a}}_{{b}}
v^{{b}}_{{\a}} \qquad
\Longleftrightarrow  \qquad
(v^{-{1}}\delta v)^{{ a}}_{{ b}} =
{1\over 4}
\Omega ^{{ (k)(l)}}(\delta )
(\Gamma _{{ (k)(l)}})^{{ a}}_{{ b}}
\qquad
\end{equation}
with the Cartan form
$\Omega ^{{ (k)(l)}}(\d)$ \p{1.14} as the parameter.

This result seems to be just evident when the definition
\p{1.1} of the harmonic variables is taken into account.

\bigskip
\bigskip

\begin{center}
{\sl \large {2.2. Twistor-like action functional \\
for $N=1$ SUPER--$P$--BRANES in D-dimensions}}
\end{center}

\bigskip
\bigskip

The proposed in \cite{bzst,bzm} action functional for super-$p$-branes
moving in space--time of any admissible \cite{m} dimension $D$ has the
following form
\begin{equation}\label{2.1}
S_{{ D,N=1,p}} = (\a ')^{-{1 \over 2}}
\int d^{{ p+1}} \xi \hskip9pt e(\xi )
\left( -  e^{\mu}_{{f}} \omega ^{{ m}}_{\mu }
u^{{ \{ f \} }}_{{ m}} + c (\a ')^{{1 \over 2}} \right) +
S^{{ W-Z}}_{{D,N=1,p}} ,
\end{equation}
\begin{eqnarray}\label{2.1a}
S^{{ W-Z}}_{{ D,N=1,p}} \equiv
(\a ')^{-{1 \over 2}} {ai \over {p+1}}
\int  d^{{p+1}}\xi \hskip9pt
\epsilon^{\mu _{{ p}}\ldots \mu _{{ 1}}\mu _{{ 0}}}
\sum^{{p+1}}_{{k=0}} [ \omega ^{{ m}_{{ p}}}_{\mu _{{ p}}}
\ldots \omega^{{ m}_{{k+1}}}_{\mu_{{ k+1}}} \times \nn \nn
\partial_{\mu_{{ k}}} x^{{ m}_{{ k}}}
\ldots \partial _{\mu _{{ 1}}}x^{{ m}_{{ 1}}} (\partial _{\mu _{{ 0}}}
\theta \Gamma _{{ m}_{{ 1}}\ldots {m}_{{ p}}}C^{-{ 1}}\theta )].
\end{eqnarray}
Here $\a '$ is the Regge slop like parameter with the dimension equal the
square of length, $c$ is a dimensionless parameter
($c (\a ')^{{1 \over 2}})$ is the
\ws{} cosmological constant) and the value of
the parameter $a$
\begin{equation}\label{2.1b}
a= \pm i^{{p(p-1)} \over 2}
(c^{{2}} \a ')^{-{p \over 2}} {{p^p} \over {p!}}
\end{equation}
is defined (up to a sign factor) by the \ks{}
requirement;
\begin{equation}\label{2.01a} \omega ^{{ m}}_{\mu } =
\partial _{\mu}x^{{ m}} -
i \partial _{\mu }\th \G ^{{m}}C^{-{1}} \th \equiv
\partial _{\mu }x^{{m}} - i \partial _{\mu }\th^{\a}
\left(\G^{{m}}C^{-{1}}\right)_{\a \b}\th^{\b} ,
\end{equation}
$x^{{m}} (m= 0,1,\ldots ,D-1)$ are the ordinary (flat) space-time coordinates
and
$\th ^{\a} (\a =1,\ldots ,2^\nu)$ are  the fermionic
(Grassmannian) coordinates of the $N=1$ $D$--dimensional superspace which
have the properties of the Majorana (Majorana-Weyl for $D=2,10$) spinors
with respect to $SO(1,D-1)$  group.

\bigskip
\bigskip
\begin{center}
{\sl \large {2.3. SUPER--$P$--BRANE equations of motion}}
\end{center}

\bigskip
\bigskip

Using the admissible variation concept \p{1.16} we can write an
arbitrary variation of the functional \p{2.1} in the form
\footnote{The use of the supersymmetric invariant variation
$ \omega ^{{ m}}= \d x^{{m}} - i \d \th \G ^{{m}}C^{-{1}} \th $
instead of $\d x^{{m}}$ simplifies the derivation of the motion equations
and symmetry transformations in the manifestly supersymmetric form.}
\begin{eqnarray}\label{2.2}
\delta S_{{ D,N=1,p}}
= (\alpha ^\prime)^{-{ 1/2}} \hskip6pt
\int d^{{p+1}}\xi \hskip6pt [ e\delta e^{\mu}_{{f}}
\{ - \omega ^{{ m}}_{\mu } u^{{ \{f\}}}_{{ m}} -
e^{{ f}}_{\mu} ( - \omega ^{{ m}}_{\nu } u^{{ \{ g \}}}_{{ m}}
e^{\nu }_{{ g}} + c (\alpha ^\prime )^{{{1/2}}} )\} \}+
\nn
\nn
+ \Omega^{{\{g\}\{f\}}}(\delta ) ee^{\mu }_{{ f}} u_{{ \{ g\}m}}
\omega ^{{ m}}_{\mu } +
\Omega ^{{ (i)\{f\}}}(\delta ) e e^{\mu }_{{ f}}
\omega ^{{ m}}_{\mu } u^{{ (i)}}_{{m}} + \hskip24pt
\nn
\nn
+ \omega ^{{ m}}(\delta ) \{ \partial _{\mu } \left(
ee^{\mu }_{{ f}} u^{{ \{ f \}}}_{{ m}} \right) +
ipa \epsilon^{\mu_{{ p}}\ldots \mu _{{ 1}}\mu _{{ 0}}}
\omega ^{{ m}_{{p}}}_{\mu _{{ p}}}
\ldots \omega ^{{ m}_{{ 2}}}_{\mu _{{ 2}}}
\partial _{\mu _{{ 1}}}\theta
\Gamma _{{ mm}_{{ 2}}\ldots { m}_{{ p}}}C^{-{ 1}}
\partial _{\mu _{{ 0}}}\theta  \} + \nn
\nn
+ 2i \{e e^{\mu }_{{ f}} (\partial _{\mu }\theta
\Gamma _{{ m}}C^{-{ 1}})_{\alpha }
u^{{ \{ f \} }}_{{ m}} +
a \epsilon ^{\mu _{{ p}}\ldots \mu _{{ 1}}\mu _{{ 0}}}
\omega ^{{ m}_{{ p}}}_{\mu _{{ p}}}\ldots
\omega ^{{ m}_{{ 1}}}_{\mu _{{ 1}}}
(\partial _{\mu _{{ 0}}}\theta
\Gamma _{{ m}_{{ 1}}\ldots { m}_{{ p}}}C^{-{ 1}})_{\alpha } \}
\delta \theta ^{\alpha } ] ,
\end{eqnarray}

Hence, the motion equations for the discussed super-$p$-brane
formulation have the following form

\begin{equation}\label{2.2a}
\omega^{{ m}}_{\mu }
u^{{ \{f\}}}_{{ m}} = {c \over p} (\a^\prime )^{{+ {1 \over 2}}}
e^{{ f}}_{\mu } ,
\end{equation}

\begin{equation}\label{2.2b}
\omega ^{{ m}}_{\mu } u^{{ (i)}}_{{ m}} = 0 ,
\end{equation}

\begin{equation}\label{2.2c}
\partial _{\mu }
\left(
ee^{\mu }_{{ f}}u^{{ \{f\}}_{{ m}}}\right)  +
ipa \epsilon ^{\mu _{{ p}}\ldots \mu _{{ 1}}\mu _{{ 0}}}
\omega ^{{m}_{{ p}}}_{\mu _{{ p}}}\ldots
\omega ^{{ m}_{{ 2}}}_{\mu _{{ 2}}} \partial_{\mu _{{ 1}}}
\theta \Gamma _{{ mm}_{{ 2}}\ldots{m}_{{ p}}}
C^{-{ 1}} \partial _{\mu _{{ 0}}}\theta  = 0
\end{equation}

\begin{equation}\label{2.2d}
e^{\mu }_{{ f}} \partial _{\mu }\theta ^{\beta } v^{{ b}}_{\beta }
\left(\Gamma ^{\{{ f\}}}C^{-{ 1}}\right)_{{ bc}}
\left(\delta ^{{ c}}_{{ a}}
\pm (\Gamma ^\prime )^{{ c}}_{{ a}} \right) = 0
\end{equation}
In the derivation of Eq.\p{2.2d} the explicit value of parameter $a$
was substituted and the following identity was used
\begin{equation}\label{2.7a}
\epsilon ^{f_{p} \ldots f_{ 1} f}
(\Gamma _{\{f_{1}\} \ldots \{f_{ p} \}} C^{-{ 1}})_{{ ba}} =
- i^{{{p(p-1) \over 2}}} p!  (\Gamma^{\{ f \}} C^{-{ 1}}
\Gamma^\prime)^{c}_{a}
\end{equation}
where the matrix $\Gamma^{\prime}$ is defined by the relation
\begin{equation}\label{2.7b}
(\Gamma ^\prime )^c_a \equiv {1 \over {(p+1)!}}
i^{{{p(p-1) \over 2}}}
\epsilon ^{f_p \ldots f_1 f_0}
\left( C\Gamma _{\{f_0 \} \ldots \{f_p\} } C^{-1}\right)^c_a
\end{equation}
and is the square root from the unity matrix
\begin{equation}\label{2.7b1}
(\Gamma ^\prime )^{{ 2}} = I \hskip64pt
\end{equation}
So,
$
{1 \over 2} (1 \pm \Gamma^{\prime})
$
are the projectors, which are related to
the known ones
$
{1 \over 2} (1 \pm \Gamma)
$
building from the matrix \cite{m1}
\begin{equation}\label{2.7c}
(\Gamma)^{\a}_{\b}
\equiv
{ i^{{{p(p-1) \over 2}}} \over {(p+1)!}}
\left({ p \over {c (\a^\prime)^{{1/2}} } }
\right)^{{p+1}}
{1 \over e}
\epsilon ^{\mu_p \ldots \mu_1 \mu_0}
\omega^{m_{{p}}}_{\mu_{{p}}} \ldots \omega^{m_{{1}}}_{\mu_{{1}}}
\omega^{m_{{0}}}_{\mu_{{0}}} \left( C\Gamma _{\{m_{0} \} \ldots
\{m_p\} } C^{-1}\right)^{\a}_{\b}
\end{equation}
on the mass shell
{\footnote{To reduce the expression \p{2.7c} for the projector $\Gamma$ to
the form presented in \cite{m1} the value of the dimensionless constant
$c$ should be fixed to be $p i^{{1/(p+1)}}$}}.

However, ${1 \over 2}(1 \pm \Gamma ^\prime)$
has the projection
properties {\bf {off--shell}} in distinction with
${1 \over 2}(1 \pm \Gamma)$.

This is one of
the advantages of the twistor--like approach.

After the exclusion of the Lorentz-harmonic variables using
\p{2.2a}, \p{2.2b}, the equations of motion \p{2.2c}, \p{2.2d}
coincide with the standard ones \cite{gsw,m1,m}. This proves the
classical equivalence of the discussed formulation with standard
one on the level of motion equations. The proof on the level of
action functionals is presented in the next subsection.

However, equations of the Lorentz-harmonic formulation have the simpler
form.  So, Eq. \p{2.2c} has the $\s$-model-like form.
The simplicity of Eq. \p{2.2d} is evident when the
$ SO(1,9)_L \times [SO(1,p) \times SO(D-p-1)]_R$--invariant splitting of
the harmonic matrix
\begin{equation}\label{split}
v^{a}_{\a} = \left(
v^q_{\a A} , v_{q \a \dot A} \right)
\end{equation}
is used. Here $q$ is the
spinor index of $SO(1,p)$, $A$ and $\dot A$ are the indices of some (may
be, the coinciding) spinor representations of $SO(1,D-p-1)$ group. Because
of the projection character of the matrices
${1 \over 2}(1 \pm \Gamma^\prime)$ there exists a
representation where the matrices
$
\left(\Gamma ^{\{{ f\}}}C^{-{ 1}}\right)_{{ bc}}
\left(\delta ^{{ c}}_{{ a}} \pm (\Gamma ^\prime )^{{ c}}_{{ a}} \right)
$
have the diagonal form with only nonvanishing components

\begin{eqnarray}\label{rep}
\left((\Gamma ^{{ \{ f\}}} C^{-{ 1}})
(I \pm \Gamma^\prime )\right)^{qp}_{ {\dot A} {\dot B}}
\propto
\delta_{{\dot A \dot B}} (\epsilon \gamma^{{f}})^{{qp}}
\nn \nn
\left((\Gamma ^{{ \{ f\}}} C^{-{ 1}})
(I \mp \Gamma^\prime )\right)_{ {qA} {pB}}
\propto
\delta_{{A B}} (\gamma^{{f}} \epsilon^{{-1}} )_{{qp}}
\end{eqnarray}
where $\gamma^{{f}}$ and $\epsilon$ are $d=(p+1)$--dimensional
$\g$--matrix and charge--conjugation matrix respectively.

In such representation Eq.\p{2.2d} acquires the following simple
particle--like form
\begin{equation}\label{2.2d1}
e^{\mu }_{{ f}} \partial_{\mu }\theta ^{\beta }
v_{{\beta q \dot A}} (\epsilon \gamma^{{f}} )^{{qp}}
= 0
\end{equation}

Note that Eq.\p{2.2d1} can be presented in the form of the Dirac
equation
\begin{equation}\label{2.2d2}
(\epsilon \gamma^{{f}})^{{pq}} D_{{f}} \theta_{{q \dot A}} = 0
\end{equation}
for the variable
$\theta_{{q \dot A}} = \theta^{\b} v_{{\beta q \dot A}}$
which appears  as a covariant piece of the
Lorentz invariant Grassmann field
$\theta^{{a}} = \theta^{\b} v^{{a}}_{{\beta}} =
 \left( \theta^{{p}}_{{A}}, \theta_{{p \dot A}} \right)$.
 The
covariant derivative $D_{{f}}$ involved into Eq.\p{2.2d2} is
defined by the relation
\begin{equation}\label{2.2d3}
D_{{f}} \theta^{{a}} \equiv
e^{\mu }_{{ f}}
\left(\partial_{\mu }\theta^{{a}} -
{1 \over 4} \Omega^{{(k)(l)}}_{{\mu}}
\theta^{{b}} (\Gamma_{{(k)(l)}})^{{a}}_{{b}}
\right) \qquad
\end{equation}
with $\Omega^{{(k)(l)}}_{{\mu}} $ to be the components of the $SO(1,D-1)$
covariant Cartan form \p{1.14}

 \bigskip
 \bigskip

\begin{center}
{\sl \large {2.4. Classical equivalence with standard formulations}}
\end{center}

\bigskip
\bigskip

Eqs.\p{2.2a} and \p{2.2b} means that the vectors $u^{ \{ f \} }_{m}$ are
tangent to the \ws{} and the vectors $u^{(i)}_{m}$ are orthogonal to the
\ws
\begin{equation}\label{2.2a1}
\omega^{{ m}}_{\mu } =
{c \over p} (\a^\prime )^{{+ {1 \over 2}}}
e^{{ f}}_{\mu } u^{ { \{f\}} { m}} ,
\end{equation}
{}From the other hand, Eq.\p{2.2a} results in the
fact that the first term in the Lagrangian \p{2.1} is proportional to the
second one
\begin{equation}\label{2.2a2}
e e^{\mu}_{{f}} \omega^{{ m}}_{\mu } u^{ { \{f\}}}_{{m}} =
 {1 \over p} e c (\a^\prime )^{{+ {1 \over 2}}} ,
\end{equation}
Further, take into account,
that $e \equiv det(e^{{f}}_{{\mu}})$ may be rewritten in terms of
induced metric $g_{{\mu \nu}}$
$$
g_{{\mu}{\nu}} \equiv e^{{f}}_{{\mu}} e_{{f}{\nu}} =
{{p^{2}} \over {c^{2} \a^{\prime}}}
\omega^{m}_{\mu} \omega_{{m}{\nu}}
\qquad
$$
in the form
$$
e \equiv det(e^{{f}}_{{\mu}}) \equiv
det^{{1/2}}(g_{{\mu}{\nu}})
\qquad
$$
Hence,
\begin{equation}\label{2.2a3}
e =
{{p^{{2(p+1)}}} \over { c^{{2(p+1)}} {(\a^\prime)^{p+1}} } }
det^{{1/2}}(\omega^{m}_{\mu} \omega_{{m}{\nu}})  \qquad
\end{equation}

Substituting Eq.\p{2.2a3} into the action functional \p{2.1} we
get the standard action in the Dirac--Nambu--Goto--like form
\cite{m1} (up to numarical constant)
$$ S_{{{Dirac-Nambu-Goto}}} =
{{p^{{2(p+1)}} (p-1)} \over {{c^{{2(p+1)}} (\a^\prime)^{{p+1}}}}}
\int d^{{p+1}} \xi
det^{{1/2}}(\omega^{m}_{\mu} \omega_{{m}{\nu}}) + S_{{W-Z}}
$$
where $S_{{W-Z}}$ has the form \p{2.1a}.

This concludes the proof of the classical equivalence of the
discussed twistor--like formulation of super--$p$--branes with
the standard ones \cite{m1}.

\bigskip
\bigskip

{\sl \large {2.5. {\bf Irreducible} \ks{} transformations}}

\bigskip
\bigskip

To derive the form of the \ks{} transformations it is useful to
rewrite the action variation by extracting the blocks
proportional to the  left hand sides of motion equations \p{2.2a}
-- \p{2.2c}. First of all we transform the terms containing the
auxiliary fields variations $\d e$ and $\d v \propto \Omega (\d)$
(see \p{1.16}). As a result the first two terms in Eq. \p{2.2}
turn into
\begin{eqnarray}\label{2.3}
\int  d^{{ p+1}}\xi [ \{ - e\delta e^{\mu }_{{ f}} + e\delta
e^{\nu }_{{ g}}e^{{ g}}_{\nu } e^{\mu }_{{ f}} + e e^{{ g\mu
}}\Omega _{{ gf}}(\delta ) \}
\{ (\alpha ^\prime )^{-{ 1/2}} \omega ^{{ m}}_{\mu }
u^{{ \{f\}}}_{{ m}} - {1\over p} e^{{ f}}_{\mu } \} +
\nn \nn
\Omega ^{{ (i)\{f\}}}(\delta ) (\alpha ^\prime )^{-{ 1/2}}
e e^{\mu }_{{ f}} \omega ^{{ m}}_{\mu } u^{{ (i)}}_{{ m}} ]
\hskip36pt \qquad
\end{eqnarray}
To transform in the similar way the terms involving  $\om^m (\d)$ and
$\d \Th$ the following relations should be used
\begin{eqnarray}\label{2.4}
\epsilon ^{\mu _{{ p}}\ldots \mu _{{ 1}}\mu _{{ 0}}}
\omega ^{{ m}_{{ p}}}_{\mu _{{ p}}}\ldots
\omega ^{{m}_{{ 1}}}_{\mu _{{ 1}}}
(\partial _{\mu _{{ 0}}}\theta
\Gamma _{{ m}_{{ 1}}\ldots { m}_{{ p}}}C^{-{ 1}})_{\alpha } =
\hskip24pt \nn \nn
=\epsilon ^{\mu _{{ p}}\ldots \mu _{{ 1}}\mu _{{ 0}}}
\{ \omega ^{{ m}_{{ p}}}_{\mu _{{ p}}}
u^{{ (n}_{{ p}}{ )}}_{{ m}_{{ p}}}\}\ldots
\{ \omega ^{{ m}_{{ 1}}}_{\mu _{{ 1}}}
u^{{ (n}_{{ 1}}{ )}}_{{ m}_{{ 1}}}\}
\partial _{\mu _{{ 0}}}\theta ^{\gamma }
 v^{{ b}}_{\gamma }
 (\Gamma _{{ (n}_{{ 1}}{ )\ldots (n}_{{ p}}{ )}}C^{-{ 1}})_{{ ba}}
 v^{{ a}}_{\alpha } =
 \nn \nn
 = \left( c (\alpha ^\prime )^{{ 1/2}}/p \right)^{{ p}}
 e e^{\mu }_{{ f}} \partial _{\mu }
 \theta ^{\gamma } v^{{ b}}_{\gamma }
 \epsilon ^{{ f}_{{ p}}\ldots { f}_{{ 1}}{ f}}
 (\Gamma _{{ \{f}_{{ 1}}{ \} \ldots \{f}_{{ p}}{ \}}} C^{-{ 1}})_{{ ba}}
  v^{{ a}}_{\alpha } +
\nn
\nn
+ \{ (\alpha ^\prime )^{-{ 1/2}} \omega ^{{ m}}_{\mu }
u^{{ \{f\}}}_{{ m}} -
{c\over p} e^{{ f}}_{\mu } \} H^{\mu }_{{
\{f\}a}} v^{{ a}}_{\alpha } + \omega ^{{ m}}_{\mu } u^{{ (i)}}_{{ m}}
G^{\mu { i}}_{{ a}} v^{{ a}}_{\alpha } ,
\end{eqnarray}
In Eq.\p{2.4} $H$ and $G$ are defined as follows
\begin{eqnarray}\label{2.5}
H^{\mu }_{{ \{f\}a}} =
- \sum^{{ p}}_{{ k=1}}
\left( {{c (\alpha ^\prime )^{{ 1/2}}} \over p} \right)^{p-k}
\epsilon ^{\mu _{{ p}}\ldots \mu _{{ 1}}\mu }
 e^{{ f}_{{ p}}}_{\mu _{{ p}}} \ldots e^{{ f}_{{ k+1}}}_{\mu _{{ k+1}}}
 \times
 \nn \nn
 \{(\alpha^\prime )^{-{ 1/2}} \omega ^{{ m}}_{\mu _{{ k}}}
 u^{{ \{f}_{{ k}}{\}}}_{{ m}} -
 {c \over p} e^{{ f}_{{ k}}}_{\mu _{{ k}}} \}
\ldots \{(\alpha ^\prime )^{-{ 1/2}} \omega ^{{ m}}_{\mu _{{ 2}}}
u^{{ \{f}_{{ 2}}{ \} }}_{{ m}} -
{c\over p} e^{{ f}_{{ 2}}}_{\mu _{{ 2}}} \}
\times
\nn \nn
\partial _{\mu _{{ 1}}}\theta ^{\beta } v^{{ b}}_{\beta }
(\Gamma _{{ \{ f\}\{f}_{{ 2}}{ \}
\ldots \{f}_{{ p}}{ \} }} C^{-{ 1}})_{{ ba}} , \qquad
\end{eqnarray}
\begin{eqnarray}\label{2.6}
G^{\mu { i}}_{{ a}} = \sum^{{ p}}_{{ k=1}}(-1)^{{ k}}
\epsilon^{\mu _{{ p}}\ldots \mu _{{ 1}}\mu }
\omega^{{ m}}_{\mu _{{ p}}} u^{{ \{f}_{{ p}}{ \}}}_{{ m}}
\omega^{{ m}}_{\mu _{{ k+1}}}u^{{\{f}_{{ k+1}}{ \}}}_{{ m}}
\omega ^{{ m}}_{\mu _{{ k}}} u^{{ (i}_{{ k}}{ )}}_{{ m}}
\omega ^{{ m}}_{\mu _{{ 2}}} u^{{ (i}_{{ 2}}{ )}}_{{ m}} \times
\nn \nn
(\Gamma _{{ (i)(i}_{{ 2}}{ )...(i}_{{ k}}{ )}}
\Gamma _{{ \{f}_{{ k+1}}{\}\ldots \{f}_{{ p}}{\}}}
C^{-{ 1}})_{{ ba}}
\end{eqnarray}

Substituting Eqs. \p{2.3}, \p{2.4}, \p{2.7a} and the explicit value of
constant $a$ \p{2.1b} into Eq.\p{2.2} we get the following expression for
action variation
\begin{eqnarray}\label{2.8}
\d S_{D,N=1,p} = (\a^\prime)^{-1/2}
\int  d^{{ p+1}}\xi [e \{ - \d e^{\mu }_{f} +
\d e^{\nu}_{g} e^{g}_{\nu} e^{\mu}_{f}
+ e^{g\mu} \Omega _{gf}(\d) -
2ia(\a^\prime)^{{ 1/2}} e^{-{ 1}}
H^{\mu }_{\{ f \}a} v^{{ a}}_{\alpha } \delta \th^{\a} \}
\times
\nn \nn
\{ \omega ^{{ m}}_{\mu } u^{\{ f \}}_{m} -
{c \over p} (\a^\prime)^{+1/2} e^{f}_{\mu} \} + \hskip48pt
\nn \nn
+( e e^{\mu}_{f} \Omega ^{(i)\{ f \}}(\d) -
2ia G^{\mu i}_{a} v^{a}_{\a} \delta \theta ^{\alpha } ) \omega
^{{ m}}_{\mu } u^{(i)}_{m} +  \hskip42pt
\nn \nn
+ \omega^m (\d) \{ \partial_{\mu }
\left( ee^{\mu}_f u^{{ \{f\}}}_{{ m}} \right) + {\it
i}p{\it a} \epsilon ^{\mu_p \ldots \mu _1 \mu _0} \omega^{m_p}_{\mu_p}
\ldots \omega ^{m_2}_{\mu_2} \partial _{\mu _1} \theta  \Gamma _{m m_2
\ldots m_p}C^{-1} \partial_{\mu_0} \theta  \} + \hskip12pt
\nn \nn
+ 2i e e^{\mu}_f \partial_{\mu}\th^{\b} v^b_{\b}
\left(\Gamma^{\{f\}}C^{-1} \right)_{bc}
\left(\d^c_a
-i^{{{p(p-1)} \over {2}}}
a ({{c(\a^\prime)^1/2} \over p})^p p!
(\Gamma ^\prime )^c_a \right) v^{{a}}_{\alpha } \d \th^{\a}  \qquad
\end{eqnarray}
The demand that $\d S_{D,N=1,p} $ \p{2.8} vanish permits to find the
explicit form of the \ks{} transformations and the values \p{2.1b}
of the coefficient $a$ of the Wess-Zumino term \p{2.1a} for which the
action \p{2.1} is invariant under these transformations.

Indeed for this (and only for this) values of parameter $a$ the matrix
$$
\left(\d^c_a
-i^{{{p(p-1)} \over {2}}}
a ({{c(\a^\prime)^1/2} \over p})^p p!
(\Gamma ^\prime )^c_a \right) ,
$$
involved into a second brackets, becomes the projection operator
$$
\left(\d^c_a \mp (\Gamma ^\prime )^c_a \right)
$$
The same projector appears in the motion equation \p{2.2d}. Hence, the
multiplication of \p{2.2d} on the second projector
$
\left(\d^c_a \mp (\Gamma ^\prime )^c_a \right)
$
from the right hand side results in the identity. In accordance with the
Second Noether theorem, this means the presence of symmetry in the theory,
which is just the \ks.

Hence, the general form of the \ks{} transformations for
$D$-dimensional super--$p$--branes in twistor-like Lorentz--harmonic
formulation is characterized by the relation

\begin{equation}\label{2.10a}
v^{{a}}_{\a} \d \th^{\a} =
\left( \d^{{ a}}_{{ b}} \mp
(\Gamma^\prime)^{{ a}}_{{ b}} \right) \e^{{ b}}
\end{equation}
These transformations involves the \ks{} parameter $\e^b$ only in the
contraction with projector
$
\left( \d^{{ a}}_{{ b}} \mp
(\Gamma^\prime)^{{ a}}_{{ b}} \right) \e^{{ b}}
$
which kills half of $2^\nu$ components of $\e^{{ b}}$. Hence, the \ks{}
has only $2^{\nu-1}$ parameters.

Indeed, in the representation \p{rep} Eq.\p{2.10a} simplifies essentially
and takes the form

\begin{equation}\label{2.11}
v^{{a}}_{\a}
\d \th^{\a} = \e^q_{A} v^{\a}_{{q A}}
\end{equation}
where $\e^q_{A}$ is the $2^{\nu -{ 1}}$ -component parameter of
the {\sl irreducible} \ks{} and $v^{\a}_{{q A}}$ is the component
of inverse harmonic matrix (see \p{split})
\begin{equation}\label{split1}
(v^{-1})^{\a}_{a} \equiv
v^{\a}_{a} = \left(\matrix{v^q_{\a A} \cr
v_{q \a \dot A}\cr} \right)
\end{equation}

The transformation rule for $x^m$ is defined by the (target space)
supersymmetric condition
\begin{equation}\label{2.10b}
\omega^{{ m}}(\d) =
\d x^{{m}} - i \d \th \Gamma ^{{m}}C^{-{1}}\th = 0 \qquad
\Rightarrow
\qquad
\d x^{{ m}} =\d \th \Gamma^{{m}} C^{-{1}} \th ,
\end{equation}
and the transformations for the auxiliary fields $e^{\mu}_{{f}}$
and $v^{a}_{\a}$ are defined by the relations

\begin{equation}\label{2.10d}
\d \left( ee^{\mu}_{{f}} \right) =
ee^{\mu { g}} \Omega_{\{g\}\{f\}}(\d)
- 2ia(\a^\prime)^{{ 1/2}} H^{\mu }_{{ \{f\}a}}
v^{{a}}_{\a} \d \th^{\a}
\end{equation}

\begin{equation}\label{2.10c}
\Omega ^{{(i)\{f\}}}(\d) = 2ia(\a^\prime)^{{1/2}}e^{-{1}}
G^{\mu {i}}_{{a}} v^{{a}}_{\a} \d \th^{\a}
\end{equation}
where $H^{\mu }_{{ \{f\}a}}$ and $G^{\mu {i}}_{{a}}$ are defined by
\p{2.5}, \p{2.6}.

Of course, the $SO(1,p)$ Cartan form
$\Omega ^{\{f\}\{g\}}(\d)$ is not determined by the requirement of
the action invariance.  This fact means  the $SO(1,p)$ gauge
invariance of  the discussed  action.  For the \ks{}
transformations we may set

\begin{equation}\label{2.10e}
\Omega ^{\{f\}\{g\}}(\d) = 0
\end{equation}

Hence, we present the action functional, the form of the {\bf
irreducible} \ks{} transformations and the equations of motion
for the twistor--like formulation of any $N=1$ super--$p$--brane
moving in space--time of arbitrary admissible (see \cite{m1})
dimension. In the next section we discuss the most interesting
cases of Heterotic superstring in D=10 and supermembrane in D=11.

\section{{\it Example 1}: $D=10$ Heterotic superstring}

The action functional for the heterotic superstring in the
twistor-like Lorentz-harmonic formulation \p{2.1} may be
specified as follows \footnote{Heterotic fermion term is omitted
here. This form of the action follows from the twistor--like
formulation for $D=10, N=IIB$ Green--Schwarz superstring
\cite{bzst} after the trivial reduction to the $N=1$ case:
$\theta^{{2}}= 0$.}

\begin{equation}\label{3.1} S_{{ 10,N=1,1}} = S_{{ 1}} +
S_{{ W-Z}},
\end{equation}
\begin{eqnarray}\label{3.1a}
S_{{ 1}} = (\alpha ')^{-{ 1/2}}
\int d \tau d \sigma  e {\Large (} c(\alpha ')^{{+ {1 \over 2}}}
- (e^{\mu { [+2]}} u^{{ [-2]}}_{{ m}}+ e^{\mu { [-2]}} u^{{
[+2]}}_{{ m}}) \omega ^{{ m}}_{\mu } {\Large )},
\nn \nn
\equiv
(\alpha ')^{-{ 1/2}} \int  d \tau d \sigma  e (
c (\alpha ')^{{+ 1/2}} - {{{ 1}\over { 16}}} e^{\mu { [+2]}}
\omega ^{{ m}}_{\mu } \left( v^{-}_{{\dot A}}
\tilde{\sigma} _{{ m}}
v^{-}_{{\dot A}} \right) -
\nn
- {{{ 1}\over { 16}}} e^{\mu {
[-2]}} \omega ^{{ m}}_{\mu } (v^{+}_{{ A}} \tilde{\sigma}_{{ m}}
v^{+}_{{ A}} ) )
\end{eqnarray}

\begin{equation}\label{3.1b}
S_{{ W-Z}} \equiv  -(c\alpha ')^{-{ 1}} \int  d \tau d \sigma
\epsilon ^{\mu \nu }  i \omega^{{ m}}_{\mu }
\left( \partial _{\nu }\theta \sigma _{{ m}}\theta
\right)
\end{equation}
Here $\omega ^{{ m}}_{\mu }$ is defined by the relation \p{2.01a} with
evident replacements of the $\Gamma$--matrices by the symmetric
$16\times 16$ Pauli matrices for $D=10$ space-time
$\sigma ^{{ m}}_{\alpha \beta }$ (see \cite{nis}  for the notations):
\begin{equation}\label{3.2}
\omega ^{{ m}}_{\mu }= \partial _{\mu} x^{{ m}}
- i \partial _{\mu }\theta \sigma ^{{ m}}\theta  \equiv
\partial _{\mu }x^{{ m}} - i \partial _{\mu }\theta ^{\alpha }\sigma
^{{ m}}_{\alpha \beta }\theta ^{\beta } ,
\end{equation}
$x^{{m}} (m= 0,1,\ldots ,9)$ are the ordinary (flat) space-time coordinates
and $\theta ^{\alpha } (\alpha =1,\ldots ,16)$  are  the
Grassmannian spinor coordinates of the $D=10, N=1$ superspace.
In Eq.\p{3.1a} the basic directions tangent  to the string \ws{} are
chosen to be light--like and the  world--sheet zweinbein is
parametrized as follows
\begin{equation}\label{3.3a}
e^f_\mu = \left( {1 \over 2}  \left(e^{[+2]}_\mu + e^{[-2]}_\mu
\right) , {1 \over 2} \left(e^{[+2]}_\mu - e^{[-2]}_\mu
\right)\right)
\end{equation}
\begin{equation}\label{3.3b}
e^\mu_f = \left( {1 \over 2}  \left(e^{\mu[-2]} + e^{\mu[+2]}
\right) , {1 \over 2} \left(e^{\mu[-2]} - e^{\mu[+2]} \right)
\right) \end{equation} The orthogonality conditions
\begin{equation}\label{3.4} e^f_\mu e^\mu_g = \d^f_g, \hskip48pt
e^\mu_f e^f_\nu = \d^\mu_\nu \end{equation}
are presented in Appendix B in the light-like notations.

The light--like notations are also convenient for the
parametrization of the
components of the
composed moving repere $u^{{(l)}}_{{m}}$  \p{1.2}, \p{1.3} of the
target space
tangent to the \ws{} .
\begin{eqnarray}\label{3.5}
u^{{ (l)}}_{{ m}}\equiv
{{ 1}\over { 16}} Sp(v^{{T}}\sigma _{{ m}}v\sigma ^{{
(l)}}) \equiv  {{ 1}\over { 16}} v^{{ a}}_{\alpha } \tilde{\sigma
}^{\alpha \beta }_{{ m}} v^{{ b}}_{\beta } \sigma ^{{ (l)}}_{{
ab}}
\equiv  (u^{{ (0)}}_{{ m}},u^{{ (1)}}_{{ m}},
\ldots,u^{{ (9)}}_{{ m}}) \equiv  (u^{{ \{f\}}}_{{ m}}, u^{{
(i)}}_{{ m}}) ,
\nn \nn
u^{{ \{f\}}}_{{ m}}= (u^{{ (0)}}_{{ m}},
u^{{ (9)}}_{{ m}}) = ( {{ 1}\over { 2}}(u^{{ [+2]}}_{{ m}}{ +
u}^{{ [-2]}}_{{ m}}) , {{ 1}\over { 2}}(u^{{ [+2]}}_{{ m}}- u^{{
[-2]}}_{{ m}}) ),
\end{eqnarray}

To get the simple expressions for composed moving frame vectors
\p{3.5} we will use the following $SO(1,9)_{Global} \times
\left( SO(1,1) \times SO(8) \right)_{Local}$ invariant splitting
of the spinor moving frame matrix variable (harmonic matrix)
$v^{{a}}_{{\a}}$ \p{1.1}
\begin{equation}\label{3.6}
v^{{ a}}_{\alpha } =
\left(
v^{+}_{\alpha {A}}, v_{\alpha {\dot A}}^{{-}}
\right)
\in \hbox{{\it Spin}}(1,9)
\end{equation}

The vectors $u^{{ [\pm 2]}}_{{ m}}, u^{{ (i)}}_{{ m}}$ are
defined in terms of
the Lorentz harmonic variables $v^{+}_{\alpha
{A}}, v_{\alpha {\dot A}}^{{-}}$ by the relations
\begin{equation}\label{3.7a}
u^{{ [+2]}}_{{ m}} = {{ 1}\over {
8}} (v^{+}_{{ A}} \tilde{\sigma }_{{ m}} v^{+}_{{ A}}) \equiv
{{ 1}\over { 8}} v^{+}_{\alpha { A}} \tilde{\sigma }^{\alpha
\beta }_{{ m}} v^{+}_{\beta { A}} ,
\end{equation}

\begin{equation}\label{3.7b}
u^{{ [-2]}}_{{ m}} = {{ 1}\over { 8}} (v^{{-}}_{\dot A}
\tilde{\sigma} _{{m}}
v^{{-}}_{{\dot A}}) ,
\end{equation}

\begin{equation}\label{3.7c}
 u^{{ (i)}}_{{ m}}= {{ 1}\over {8}}
\left( v^{+}_{{ A}} \tilde{\sigma }_{{ m}}
\gamma ^{{i}}_{{ A}{\dot A}}
v^{{-}}_{{\dot A}}) ,
\right) ,
\end{equation}
The contracted SO(1,9) spinor indices in Eqs.\p{3.7b}, \p{3.7c}
and in the following formulas are omitted.

The harmonicity conditions \p{1.1a} which realize the statement \p{1.1}
take the following form  in  the discussed case
\footnote{ Such  form of the harmonicity conditions for D=10 space-time
have been proposed in the papers \cite{ghs,gds} where superparticle case
have been discussed; the conditions \p{1.1} for $SO(10)$ group had been
discussed in the earlier work \cite{zup} and used for the discussion of
the twistor transform for the
 superfields in the recent work \cite{ght}.}

\begin{equation}\label{3.8a} \Xi _{{ m}_{{1}}\ldots { m}_{{ 4}}} =
u^{{ m(n)}} \eta _{{ (n)(l)}}
\Xi ^{{ (l)}}_{{m}_{{ 1}}\ldots { m}_{{ 4}}{ m}} = 0 ,
\end{equation}
\begin{equation}\label{3.8b}
\Xi _{{ 0}}
\equiv  u^{{ [-2]}}_{{ m}} u^{{ m[+2]}} - 2 = 0 ,
\end{equation}
The expressions
\begin{equation}\label{3.9}
\Xi ^{{ (n)}}_{{ m}_{{ 1}}\ldots{ m}_{{ 5}}}
\equiv  Sp(v^{{ T}} \tilde{\sigma }_{{ m}_{{ 1}}{\bf ...}{ m}_{{ 5}}}
v\sigma ^{{ (n)}})\equiv  v^{{ a}}_{\alpha }
(\tilde{\sigma }_{{ m}_{{ 1}}{\bf ...}{ m}_{{ 5}}})^{\alpha \beta }
v^{{ b}}_{\beta } (\sigma ^{{ (n)}})_{{ ab}} = 0
\end{equation}
vanish as the consequence of Eqs. \p{3.8a}. The last expression of  the
type \p{1.15}  vanishes  identically  because  of  the   antisymmetric
property  of  the  matrix
$(\tilde{\sigma }_{{ m}_{{ 1}} \ldots {m}_{{3}}})^{\alpha \beta }$
under   the   spinor   index permutations.

The repere orthogonality conditions are  satisfied  here as  the
common consequence of the expressions \p{3.7a}--\p{3.7c}, the
conditions \p{3.8a} and the famous identity \p{1.03}
The normalization conditions for the composed repere \p{3.7a}--\p{3.7c}
are satisfied due to the harmonicity conditions \p{3.8a}, \p{3.8b}
and  due  to  the identity \p{1.03}. The useful representation for $D=10$
$\s$--matrices is given  in the Appendix C.

Now we specify the general form of the equations of motion  \p{2.2a}--
\p{2.2d} and irreducible \ks--transformations \p{2.10a}--\p{2.11}  for
$D=10$ Heterotic string in
the twistor-like Lorentz-harmonic formulation.

The equation of motion for auxiliary fields has the form

\begin{equation}\label{3.11}
\omega ^{{ m}}_{\mu } u^{{ [\pm 2]}}_{{ m}} =
c (\alpha ')^{{ 1/2}}
e^{{ [\pm 2]}}_{\mu },
\end{equation}

\begin{equation}\label{3.15}
\omega ^{{ m}}_{\mu } u^{{ (i)}}_{{ m}} = 0
\end{equation}

Thus the light--like vectors $u^{{ m[\pm 2]}}$ are tangent to the
superstring world-sheet on the shell, defined by the motion  equations.
Contrary, the  vectors $u^{{ m(i)}}$ are orthogonal to the world-sheet on
this shell.

Using Eqs.\p{3.11}, \p{3.15}, the  classical  equivalence  of
the discussed $D=10$ superstring formulation with the  standard  Green-
Schwarz one \cite{gsw} can be justified easily (see \cite{bzst} for
$N=2$ Green--Schwarz superstring case).

The equations of motion for the $x^{{ m}}(\xi )$ and $\theta (\xi )$
fields:
${{\delta {\Large S}} \over {\delta x^{{ m}}(\xi )}} = 0$ and
${{\delta S}\over {\delta \theta ^{\alpha { I}}(\xi )}} = 0$
have the form
\begin{equation}\label{3.18}
\partial _{\mu }(e \sum^{}_{\pm } (e^{\mu {
[\pm 2]}} u^{{ [\mp 2]}}_{{ m}})) - 2i\epsilon ^{\mu \nu }\partial _{\mu
}\theta \sigma _{{ m}}\partial _{\nu }\theta /c(\alpha ')^{{ 1/2}} = 0 ,
\end{equation}
\begin{equation}\label{3.20}
(\partial _{\mu }\theta \sigma ^{{ m}})_{\alpha }
(\sum^{}_{\pm } e^{\mu { [\pm 2]}} u^{{ m[\mp 2]}} -
2 \epsilon ^{\mu \nu } \omega ^{{ m}}_{\nu }) = 0 ,
\end{equation}
which may be reduced to the standard one
\cite{gsw} in a same manner, as have been done for the general case.
Excluding the fields $\omega ^{{ m}}_{\nu }$ from  the
equation  \p{3.20},  we derive the following particle--like form
of the equation for grassmannian field $\theta$

\begin{equation}\label{3.23}
e^{\mu { [-2]}}\partial _{\mu }\theta ^{\alpha }
v^{{ +}}_{\alpha { A}} = 0 ,
\end{equation}

\section{Example 2: Supermembranes in $D=11$}

In this section we shall  use  the  following  $SO(1,2)\times SO(8)$
invariant representation  for   the   charge   conjugation   matrix and
the $\gamma $-matrices in $D=11$.
\begin{eqnarray}\label{4.1}
C^{{ ab}}= - C^{{ ba}}= {\it diag} \left(\epsilon ^{{ \hat{a}
\hat{b}}}\delta _{{ AB}},-\epsilon _{{ \hat{a}\hat{b}}}\delta
_{\dot{A}\dot{B}}\right), \nn \nn C^{-{ 1}}_{{ ab}} =\hbox{ {\it diag}}
\left( \epsilon _{{ \hat{a}\hat{b}}} \delta _{{AB}} , -\epsilon ^{{
\hat{a}\hat{b}}}\delta _{\dot{A}\dot{B}} \right) , \nn \nn \Gamma ^{{
(m)}}\equiv \left(\Gamma ^{{ \{f\}}}, \Gamma ^{{ (i)}}\right), \nn \nn
\Gamma ^{{ \{f\}}}\equiv
\left(\Gamma ^{{ \{0\}}}, \Gamma ^{{ \{1\}}}, \Gamma ^{{ \{2\}}}\right)
\equiv \left(\Gamma ^{{ 0}},\Gamma ^{{ 9}},\Gamma ^{{ 10}}\right) =
{\it diag} \left( \gamma ^{{ {f} \hat{b}}}_{{ \hat{a}}}\delta _{{ AB}},
-\gamma ^{{ {f} \hat{a}}}_{{ \hat{b}}} \delta _{\dot{A}\dot{B}}\right),
\nn \nn
\Gamma ^{{ (i)}} \equiv
\left(\Gamma ^{{ 1}},\ldots , \Gamma ^{{ 8}}\right) =
{\Large [}^{ 0
\hskip12pt
\epsilon _{{ \hat a \hat b}}\gamma ^{{ (i)}}_{{A}\dot{B}}}
_{-\epsilon ^{{ \hat{a}\hat{b}}}\tilde{\gamma }^{{ (i)}}_{{\dot A}{B}}
\hskip12pt 0} {\Large ]}
\end{eqnarray}
where
\begin{equation}\label{4.1b}
a = \left(^{\hat a}_{A}, _{{\hat a} {\dot A}} \right)
\end{equation}
is the composed spinor (upper) index of $SO(1,2)\times SO(8)$,
$\gamma ^{{ (i)}}$
are $d=8$ $\gamma $-matrices \cite{gsw}  which  are
similar those for $D=10$; $\tilde{\gamma }^{{ (i)}}_{{\dot A}{
B}}\equiv  \gamma ^{{ (i)}}_{{ B}{\dot A}}$,
$\gamma ^{{{f} \hat{a}}}_{{ \hat{b}}}$  are
$d=3$
$\gamma $--matrices, $\epsilon ^{{\hat{a} \hat{b}}}=- \epsilon
^{{ \hat{b}\hat{a}}}$ $(\epsilon ^{{ 12}}=-\epsilon _{{ 12}}=1)$
represents  $d=3$ charge conjugation matrix.

The Lorentz harmonics \p{1.1}, \p{1.1a} parametrize the
coset $SO(1,10)/(SO(1,2)\times  SO(8))$ and form $32 \times 32$
matrix $v^{{ {a}}}_{{\alpha }}$
\begin{equation}\label{4.1a}
v^{{ {a}}}_{{\alpha }}= \left( v^{\hat{a}}_{{\alpha }{ A}},
v_{{\alpha }{\hat{a}}{\dot A}} \right)
\end{equation}
where $\alpha =1,{\bf ...},32$ are  the  spinor  indices  of  the  group
$SO(1,10)$; $\hat{a}, \hat{b}=1,2$ belong to the spinor indices of
$SO(1,2)$ ; $A,B = 1,\ldots ,8$ ; ${\dot A} ,{\dot B} =1,\ldots ,8$ are
$s$- and $c$- spinor indices  of  SO(8) respectively.
The matrix \p{4.1}
takes its values in the group ${\it Spin}(1,10)$ which is  a
double-covering group for the Lorentz group $SO(1,10)$ because  it
satisfies the  following harmonicity conditions
\begin{equation}\label{4.3}
\Xi  \equiv v^{a}_{\alpha} C^{{\alpha}{\beta }}
 v^{{b}}_{\beta} - C^{{a}{b}} = 0 ,
\end{equation}
\begin{equation}\label{4.4}
\Xi ^{{ (n)}}_{{ m}_{{ 1}}{ m}_{{ 2}}} \equiv
v^{a}_{\alpha} (C\Gamma _{{ m}_{{ 1}}{ m}_{{ 2}}})^{{\alpha }{\beta }}
v^{b}_{{\beta }} \left(\Gamma ^{{ (n)}}C^{-{ 1}}\right)_{{a}{b}} = 0 ,
\end{equation}
\begin{equation}\label{4.5}
\Xi ^{{ (n)}}_{{ m}_{{ 1}}\ldots { m}_{{ 5}}} \equiv
v^{{a}}_{{\alpha }}
\left( C \Gamma _{{ m}_{{ 1}}\ldots { m}_{{ 5}}} \right)
^{{\alpha }{\beta}}
v^{{b}}_{\beta} \left( \Gamma ^{{ (n)}}C^{-{ 1}} \right)_
{{{a}{b}}} = 0 ,
\end{equation}
which exclude $496+11+462=969 (=1024-55)$ degrees of freedom.

Thus  the harmonics $v^{{ a}}_{{\alpha }{ A}} , v_{{\alpha }{\dot A}{ a}}$
describe $55 = {\sl dim}  SO(1,10) (=1024-969)$ independent    degrees
of    freedom.  Among     the latter $31= 3 + 28 = {\sl dim}  SO(1,2)+
{\sl dim}  SO(8)$ degrees of freedom are pure gauge ones due to
$SO(1,2)\times SO(8)$ local symmetry of ${\Large S}_{{ 11,1,2}}$.

The harmonicity conditions  \p{4.3} are  independent, but the relations
\p{4.4},\p{4.5} contain only 11 and 462 independent conditions on
harmonics , respectively (see \cite{bzm} for details).

The relations \p{4.3} allow to construct the matrix
inverse to $v^{{a}}_{{\alpha }}$ using the same variables
$v^{{ \hat a}}_{{\alpha }{ A}} , v_{{\alpha }{\dot A }{\hat a}}$:
\begin{equation}\label{4.7a}
\pmatrix{V^{-{ 1}}}^{{\beta }}_{{b}}= \left(
-C^{{\beta }{\alpha}} v_{{\alpha }{ {b}A}}, C^{{\beta }{\alpha }}
v^{{b}}_{{\alpha }{\dot A}} \right)
\end{equation}
Here and further the spinor indices belonging  to  SO(1,2)  group
are lifted and lowered
\begin{equation}\label{4.8}
v_{{\alpha }{ bA}}= \epsilon
_{{ ba}} v^{{ a}}_{{\alpha }{ A}} ,\qquad
v^{{ b}}_{{\alpha} {\dot A }} =
\epsilon ^{{ ba}} v_{{\alpha}{\dot A }{ a}}
\end{equation}
using the $d=3$ charge conjugation matrix.
Now it is possible to specialize the expressions \p{1.3} for  the case of
supermembrane in $D=11$ using the representation \p{4.1}
\begin{equation}\label{4.9a}
u^{{ {f}}}_{{ m}} = {1\over 32}
\left[
v^{{\hat a}}_{{ A}}
\left(C\Gamma _{{ m}}\right)
v_{{\hat b}{A}}
+
v^{{\hat a}{\dot A}} \left(C\Gamma _{{ m}}\right)
v_{{\hat b}{\dot A}}\right] (\gamma ^{f})^{{{\hat b}}}_{{\hat a}} ,
\end{equation}
\begin{equation}\label{4.9b}
u^{{ (i)}}_{{ m}} = - {1\over 16}
v^{{\hat a}}_{{ A}} \left(C\Gamma _{{ m}}\right)
v_{{\hat a}{\dot A}}
\gamma^{{ i}}_{{ A}{\dot A}}
\end{equation}
The converted SO(1,10) spinor indices are omitted in \p{4.9a},\p{4.9b}.
The use of Eqs.  \p{1.5}-\p{1.7} allows to present the expressions
\p{4.9a}, \p{4.9b} in the form

\begin{equation}\label{4.10a}
u^{{ \{f\}}}_{{ m}}
\left(
\Gamma ^{{ m}} C^{-{ 1}}
\right)_{{\alpha }{\beta }}
=
\left(
v^{{\hat a}}_{{\alpha}{ A}}
v_{{\beta}{\hat b}{A}} +
v^{{\hat a}}_{{\alpha }{\dot A}}
 v_{{\beta }{\hat b}{\dot B}} \right)
\gamma ^{{ \{ f \} {\hat b}}}_{{\hat a}} ,
\end{equation}

\begin{equation}\label{4.10b}
u^{{ (i)}}_{{ m}}
( \Gamma ^{{ m}} C^{-{ 1}})_{{\alpha }{\beta }}
= 2 v^{{\hat a}}_{{ A \{ {\alpha }}}
v_{ {\beta } \} {\hat a}{\dot A }}
\gamma ^{{ i}}_{{ A}{\dot A }}
\end{equation}

Note that the left and right parts of Eq.\p{4.10a}  are
symmetric  under permutation of  SO(1,10) spinor  indices  owing  to  the
well-known $\gamma $-matrix identities for $\gamma $-matrices in $D=11$
and $d=3$
\begin{eqnarray}\label{4.10.1}
\left(\Gamma ^{{ m}}C^{-{ 1}}\right)^{{ T}} =
\left(\Gamma ^{{ m}}C^{-{ 1}} \right) , \nn
\left(\epsilon \gamma ^{{ {f}}}\right)^{{ T}} =
\left(\epsilon \gamma ^{{ {f}}}\right)
\end{eqnarray}

The matrix \p{2.7a} for $D=11$ supermembrane ($p=2$) has the form
\begin{equation}\label{4.14}
\Gamma ^\prime  =\hbox{ diag}( \delta ^{{\hat b}}_{{\hat a}}
\delta _{{ AB}} , - \delta ^{{\hat a}}_{{\hat b}}
\delta_{{\dot A}{\dot B }})
\end{equation}

The twistor--like action for the supermembrane  in $D=11$ has the form
\p{2.1} with the Wess--Zumino term \p{2.1a} which may be
presented in the form
\begin{eqnarray}\label{4.11b}
{\Large S}^{{WZ}}_{{ 11,1,2}}=
{2\over {{\alpha ^\prime}{\sqrt {\a^\prime}} c^{2}}}
\int d \xi ^{{ 3}} \epsilon ^{\mu \nu \rho }
{\Large [}  \partial _{\mu} x^{{ m}}
{\Large (} \partial_{\nu } x^{{ n}} +
i \partial _{\nu }\theta^{{\alpha }}
(\Gamma ^{{n}} C^{-{ 1}})_{{\alpha }{\beta}} \theta ^{{\beta }}
{\Large )} - \nn
- {1\over3} \partial _{\mu } \theta
(\Gamma ^{{ m}}C^{-{ 1}})\theta \hskip6pt
\partial _{\nu} \theta (\Gamma ^{{ n}}C^{-{ 1}})\theta {\Large ]}
\hskip6pt
\partial _{\rho }\theta (\Gamma _{{ mn}}C^{-{ 1}})\theta ,
\end{eqnarray}

Equations of motion \p{2.2a}--\p{2.2d} may be specialized as
follows for $D=11$ supermembrane:
\begin{equation}\label{4.16a}
\omega ^{{ m}}_{\mu } u^{{ \{f\}}}_{{ m}} =
{c (\alpha ^\prime )^{{1/2}} \over 2}
e^{{ f}}_{\mu } ,
\end{equation}
\begin{equation}\label{4.16b}
\omega ^{{ m}}_{\mu } u^{{ (i)}}_{{ m}} = 0 ,
\end{equation}
\begin{equation}\label{4.16c}
\partial _{\mu }(ee^{\mu }_{{ f}} u^{{ \{f\} }}_{{ m}})+
{2 \over {c (\alpha^\prime )^{{ 1/2}}}}
\epsilon ^{\mu \nu \rho } e_{\mu { f}} u^{{ m\{f\}}}
\partial _{\nu } \theta \Gamma _{{ mn}}C^{-{ 1}} \partial _{\rho }\theta
= 0
\end{equation}
\begin{equation}\label{4.16d}
e^{\mu }_{{ f}} \partial_{\mu }
\theta ^{\beta }
v_{\beta {\hat b} {\dot A}}
\left(\epsilon \gamma
^{{ \{f\}}})^{{\hat b}{\hat a}}\right) = 0
\end{equation}

The expressions for the $\kappa $--symmetry transformations have the form

\begin{equation}\label{4.15a}
\delta \theta^{\alpha } v^{{ a}}_{\alpha { A}} =
\epsilon ^{{ a}}_{{ A}} ,\qquad \delta
\theta ^{\alpha } v_{\alpha {a}{\dot A}}= 0 ,\qquad
\Longleftrightarrow
\delta \theta^{\alpha } = \epsilon ^{{ a}}_{{ A}} v^{\a}_{{a}{A}}
\end{equation}
\begin{equation}\label{4.15b}
\omega ^{{ m}}(\delta ) = \delta x^{{ m}} - i \delta \theta \Gamma ^{{
m}}C^{-{ 1}}\theta  = 0\qquad \Rightarrow \qquad \delta x^{{ m}} =\delta
\theta \Gamma ^{{ m}}C^{-{ 1}}\theta  ,
\end{equation}
\begin{equation}\label{4.15c}
\Omega ^{{(i)\{f\}}}(\delta )  = 2ia(\alpha ^\prime )^{{ 1/2}}e^{-{ 1}}
G^{\mu { i}}_{{ a}} v^{{ a}}_{\alpha } \delta \theta ^{\alpha } ,
\end{equation}

\begin{equation}\label{4.15e}
\delta (ee^{\mu }_{{ f}}) =
ee^{\mu { g}} \Omega _{{ \{g\}\{f\}}}(\delta )
- 2ia(\alpha ^\prime )^{{ 1/2}} H^{\mu }_{{ \{f\}a}}
v^{{ a}}_{\alpha } \delta \theta ^{\alpha },
\end{equation}
where
\begin{eqnarray}\label{4.15d}
G^{\mu { i}}_{{ a}} v^{{ a}}_{\alpha } =
\epsilon ^{\mu _{{ 1}}\mu _{{ 2}}\mu } \omega ^{{ m}}_{\mu_{{ 1}}}
\partial _{\mu _{{ 2}}}\theta ^{\beta } \hskip6pt
[ u^{{ (i_{{ 1}})}}_{{ m}} (v^{{\hat b}}_{\beta }
\gamma _{{ii_{{1}}}}  v^{{\hat a}}_{\alpha } \epsilon _{{\hat b}{\hat a}}
- v_{\beta {\hat b}} \tilde{\gamma }_{{ ii_{{ 1}}}} v_{\alpha {\hat a}}
\epsilon ^{{\hat b}{\hat a}} ) + \nn \nn
u^{{ \{f\}}}_{{ m}} (v^{{\hat b}}_{\beta }
\gamma _{{ i}} v_{\alpha {\hat a}} +
v_{\beta {\hat b}} \tilde{\gamma }_{{ i}}
v^{{\hat a}}_{\alpha })
\end{eqnarray}
and
\begin{eqnarray}\label{4.15f}
H^{\mu }_{{ \{ f \} a}} v^{{a}}_{\alpha } =
i \epsilon ^{\mu \mu_{{ 1}} \mu _{{ 2}}}
\epsilon _{{f f_{1} f_{2}}} \omega ^{{ m}}_{\mu _{{ 1}}}
u^{{ \{ f_{{ 1}} \}}}_{{ m}}
\partial _{\mu _{{ 2}}}
\theta ^{\beta }
( v^{{\hat b}}_{\beta { A}}
(\gamma ^{{ \{ f_{{ 2}} \} }}
\epsilon ^{-{ 1}})_{{\hat b}{\hat a}}
v^{{\hat a}}_{\alpha { A}} -
v_{\beta {\hat b}{\dot A}}
(\epsilon \tilde{\gamma} ^{{ \{ f_{{ 2}} \} }})^{{\hat b}{\hat a}}
v_{\alpha {\hat a}{\dot A}})
\end{eqnarray}

As a result of $SO(1,2)$ gauge  invariance  of  the  action,
Cartan form $\Omega ^{{ \{f\}\{g\}}}(\delta )$ is not determined
and we may set
\begin{equation}\label{4.15g}
\Omega ^{{ \{f\}\{g\}}}(\delta ) = 0
\end{equation}
for the $\kappa $-symmetry transformations.

\newpage

\section{Conclusion}

Therefore, the twistor--like formulations for the all known $N=1$
super--p--brane theories \cite{t1}
are presented here. The possible
application of these formulations is the investigation of the
nonlinear equations of motion for super--$p$--brane with $p \geq 2$
using the generalized twistor methods.

Moreover, it was proved that
the \ks{} (as well as all another gauge symmetries) are present
in the irreducible form in these formulations (see Subsection 2.5).
So, the covariant Hamiltonian formalism for all set of $N=1$
super--$p$--branes can be build using the results of this paper.
(Such formalism for $D=10$ $N=IIB$ superstring have been
developed in Ref. \cite{bzst}).

We prove the classical equivalence of the discussed
super--$p$--brane formulations with the known ones. However such
equivalence can be destroyed by quantum effects as well as by the
coupling to the gauge fields.
Does this destruction indeed appears or no?
This is the interesting question for further investigation.

One more point noteworthy is connected with the recently
discovered type $IIA$ and $IIB$ super--$p$--branes previously
thought not to exist for $p \geq 2$ \cite{m2}. These type $II$
super--$p$--branes emerge as solitons of either type $IIA$ or
$IIB$ supergravity and involve additional \ws{} vector or tensor
fields. As it has been recently shown such
super--$p$--branes exist in $D=10$ only \cite{m3}. Unfortunately at the
present time supersymmetric and $\kappa$--symmetric formulations of
these type p--brane actions are absent. It could be suggested
that the twistor--like harmonic approach developed here may be
useful for solving this problem.

\section{Acknowlagements}

In conclusion we acknowledge K.S.Stelle and D.V.Volkov
 for useful conversations and helpful discussions.

One of the authors (I.A.B.) would like to thank Prof.
Mario Tonin and Prof. Paolo Pasti for the interest to the work,
the discussions of the results  and for the hospitality
in the Padova University and Padova I.N.F.N. section.

\newpage
\section{Appendix A}
\begin{center}
{\sl The Fiertz identities for $D=2,3,4(mod {~}8)$}
\end{center}

\begin{eqnarray}\label{A.1}
F^{\beta }_{\alpha } = 2^{-\nu } \delta ^{\beta }_{\alpha } Sp(F) -
2^{-(\nu +1)} (\Gamma ^{{ m}_{{ 1}}{ m}_{{ 2}}} )^{\beta }_{\alpha }
Sp(\Gamma _{{ m}_{{ 1}}{ m}_{{ 2}}} F) + \nn
+ \sum^{}_{{ k\ge 1}} B_{{ 2k}}
(\Gamma ^{{ m}_{{ 1}}{\bf ...}{ m}_{{ 2k}}} )_{\alpha \beta }
Sp(\Gamma _{{ m}_{{ 1}}{\bf ...}{ m}_{{ 2k}}}F) + \nn
+ (1- \delta _{D(\hbox{mod}8),2}) [ 2^{-\nu }
(\Gamma ^{{ m}} )^{\beta }_{\alpha } Sp(\Gamma _{{ m}} F) + \nn
+ \sum^{}_{{ k\ge 1}} B_{{ 2k+1}}
(\Gamma ^{{ m}_{{ 1}}{\bf ...}{ m}_{{ 2k+1}}} )^{\beta }_{\alpha }
Sp(\Gamma _{{ m}_{{ 1}}{\bf ...}{ m}_{{ 2k+1}}}F) ]
\end{eqnarray}
where the last term is absent for $D=2(mod~8)$,

\begin{eqnarray}\label{A.2}
U^{\alpha \beta } = 2^{-\nu } (C\Gamma ^{{ m}})^{\alpha \beta }
Sp(U\Gamma _{{ m}}C^{-{ 1}}) + \nn
+ \sum^{}_{{ k\ge 1}} B_{{ 2k+1}}
(C\Gamma ^{{ m}_{{ 1}}{\bf ...}{ m}_{{ 2k+1}}} )^{\alpha \beta }
Sp(U\Gamma _{{ m}_{{ 1}}{\bf ...}{ m}_{{ 2k+1}}}C^{-{ 1}}) \nn
+ (1 - \delta _{D(mod 8),2}) [ 2^{-\nu } C^{\alpha \beta }
Sp(UC^{-{ 1}})- \nn
- 2^{-(\nu +1)}
(C\Gamma ^{{ m}_{{ 1}}{ m}_{{ 2}}} )^{\alpha \beta }
Sp(U\Gamma _{{ m}_{{ 1}}{ m}_{{ 2}}} C^{-{ 1}}) + \nn
+ \sum^{}_{{ k\ge 2}} B_{{ 2k}}
(C\Gamma ^{{ m}_{{ 1}}{\bf ...}{ m}_{{ 2k}}} )^{\alpha \beta }
Sp(U\Gamma _{{ m}_{{ 1}}{\bf ...}{ m}_{{ 2k}}}C^{-{ 1}}) ] ,
\end{eqnarray}

\begin{eqnarray}\label{A.3}
W_{\alpha \beta } =
2^{-\nu } (\Gamma ^{{ m}}C^{-{ 1}})_{\alpha \beta }
Sp(C\Gamma _{{ m}}W) + \nn
+ \sum^{}_{{ k\ge 1}} B_{{ 2k+1}}
(\Gamma ^{{ m}_{{ 1}}{\bf ...}{ m}_{{ 2k+1}}}C^{-{ 1}})_{\alpha \beta }
Sp(C\Gamma _{{ m}_{{ 1}}{\bf ...}{ m}_{{ 2k+1}}}W) + \nn
+ (1- \delta _{D(mod 8),2})
[ 2^{-\nu } C^{-{ 1}}_{\alpha \beta } Sp(CW) -
2^{-(\nu +1)} (\Gamma ^{{ m}_{{ 1}}{ m}_{{ 2}}}
C^{-{ 1}})^{\alpha \beta }
Sp(C\Gamma _{{ m}_{{ 1}}{ m}_{{ 2}}} W) + \nn
\sum^{}_{{ k\ge 2}} B_{{ 2k}}
(\Gamma ^{{ m}_{{ 1}}{\bf ...}{ m}_{{ 2k}}}C^{-{ 1}})_{\alpha \beta }
Sp(C\Gamma _{{ m}_{{ 1}}{\bf ...}{ m}_{{ 2k}}}W)]
\end{eqnarray}

We do not need in the expressions for the numerical  coefficients $B_{{ k}}$
except for the first three ones
$$B_{{ 0}}=B_{{ 1}}=2^{-\nu },$$
$$B_{{ 2}}= - 2^{-(\nu +1)}$$

\newpage
\section{Appendix B}
\begin{center}
{\sl World-sheet repere (p=1, d=2). \\
The orthonormality conditions in "light-like" notations}
\end{center}

Repere variables of the (super)string \ws{}
$$e^{{ [\pm 2]}}_{\mu }, e^{\mu { [\pm 2]}}$$
satisfy the following conditions
\begin{equation}
e^{{ [+2]}}_{\mu } e^{\mu { [+2]}}= 0 = e^{{ [-2]}}_{\mu }e^{\mu { [-2]}} ,
\qquad
e^{{[+2]}}_{\mu } e^{\mu { [-2]}}= 2 = e^{{ [-2]}}_{\mu } e^{\mu { [+2]}} ,
\end{equation}
\begin{equation}
\epsilon ^{\mu \nu }= {{ 1}\over { 2}} e (e^{\mu { [+2]}} e^{\nu { [-2]}}-
e^{\mu { [-2]}} e^{\nu { [+2]}}) ,
\qquad
\left( \epsilon ^{{ 01}}=-\epsilon _{{ 01}}=1 \right) ,
\end{equation}
\begin{equation}
 g^{\mu \nu }= {{ 1}\over { 2}} (e^{\mu { [+2]}} e^{\nu { [-2]}}+
 e^{\mu { [-2]}} e^{\nu { [+2]}}) ,\qquad
\sqrt{-g} \equiv  e ,
\end{equation}
\begin{equation}
\delta ^{\nu }_{\mu }= {{ 1}\over { 2}} (e^{{ [+2]}}_{\mu } e^{\nu { [-2]}} { +
e}^{{ [-2]}}_{\mu } e^{\nu { [+2]}}) ,
\qquad
\epsilon _{\mu \nu } e^{\mu { [-2]}} e^{\nu { [+2]}}= 2/e ,
\end{equation}

\section{Appendix C}

The following representation for the $\sigma$-matrices should be used
for an explicit calculations in the case of 10--dimensional
space-time:
\begin{equation}
\sigma ^{{ 0}}_{{ ab}}=\hbox{ {\it diag}}(\delta _{{ AB}},
\delta_{{\dot A}{\dot B}})
= \tilde{\sigma }^{{ 0 ab}} ,
\end{equation}
\begin{equation}
\sigma ^{{ 9}}_{{ ab}}= \hbox{ {\it diag}} (\delta _{{ AB}},
-\delta _{{\dot A}{\dot B}}) = -\tilde{\sigma }^{{ 9ab}} ,
\end{equation}
\begin{equation}
\sigma ^{{ (i)}}_{{ ab}} =
\left(\matrix{0 & \gamma ^{i}_{{ A}{\dot B}}\cr
\tilde{\gamma}^{i}_{{\dot A}{ B}} & 0\cr}
\right)
= - \tilde{\sigma }^{{(i)ab}} ,
\end{equation}
\begin{equation}
\sigma ^{{ [+2]}}_{{ ab}}\equiv (\sigma
^{{ 0}}+\sigma ^{{ 9}})_{{ ab}}=\hbox{ {\it diag}}(2\delta _{{ AB}}, 0)
= -(\tilde{\sigma }^{{ 0}}- \tilde{\sigma }^{{ 9}})^{{ ab}} =
\tilde{\sigma }^{{ [-2]ab}} ,
\end{equation}
\begin{equation}\label{3.11'}
\sigma ^{{ [-2]}}_{{ ab}}\equiv
(\sigma ^{{ 0}}-\sigma ^{{ 9}})_{{ ab}}=\hbox{ {\it diag}}(0, 2\delta
_{{\dot A}{\dot B}}) = (\tilde{\sigma }^{{ 0}}+
\tilde{\sigma }^{{ 9}})^{{ ab}} = \tilde{\sigma }^{{ [+2]ab}} ,
\end{equation}
Here
$\gamma ^{i}_{{A}{\dot B}}$
are the
$\sigma$
-matrices  for $SO(8)$ group,
$
\tilde{\gamma }^{i}_{{\dot A}{B}} \equiv
\gamma^{i}_{{ B}{\dot A}}
$ .

  \end{document}